\documentclass[sigconf]{acmart}

\usepackage{tabularx}
\usepackage{booktabs}
\usepackage{tcolorbox}
\usepackage{algorithm}
\usepackage{algpseudocode}
\usepackage{multirow}
\usepackage{xcolor}

\copyrightyear{2026}
\acmYear{2026}
\setcopyright{cc}
\setcctype{by}
\acmConference[SIGIR '26]{Proceedings of the 49th International ACM SIGIR Conference on Research and Development in Information Retrieval}{July 20--24, 2026}{Melbourne, VIC, Australia}
\acmBooktitle{Proceedings of the 49th International ACM SIGIR Conference on Research and Development in Information Retrieval (SIGIR '26), July 20--24, 2026, Melbourne, VIC, Australia}
\acmDOI{10.1145/3805712.3808410}
\acmISBN{979-8-4007-2599-9/2026/07}

\begin{document}

\title{RecGPT-Mobile: On-Device Large Language Models for User Intent Understanding in Taobao Feed Recommendation}


\author{Bin Zhang}
\authornote{Both authors contributed equally to this research.}
\email{tianji.zb@taobao.com}
\author{Weipeng Huang}
\authornotemark[1]
\email{weipeng.hwp@taobao.com}
\author{Dimin Wang}
\email{dimin.wdm@taobao.com}
\author{Jialin Zhu}
\email{xiafei.zjl@taobao.com}
\affiliation{%
  \institution{Taobao \& Tmall Group of Alibaba}
  \city{HangZhou}
  \country{China}
}

\author{Yuning Jiang}
\email{mengzhu.jyn@taobao.com}
\author{Zhaode Wang}
\email{zhaode.wzd@taobao.com}
\author{Chengfei Lv}
\email{chengfei.lcf@taobao.com}
\author{Jian Wang}
\email{krod.wj@taobao.com}
\affiliation{%
  \institution{Taobao \& Tmall Group of Alibaba}
  \city{HangZhou}
  \country{China}
}

\author{Qichao Ma}
\email{maqichao.mqc@taobao.com}
\author{Li Chen}
\email{cl121469@taobao.com}
\author{Junqing Wu}
\email{junqing.wjq@taobao.com}
\author{Yipeng Yu}
\email{linxin.yyp@taobao.com}
\affiliation{%
  \institution{Taobao \& Tmall Group of Alibaba}
  \city{HangZhou}
  \country{China}
}

\renewcommand{\shortauthors}{Zhang et al.}


\begin{abstract}
Predicting a user’s next search query from recent interaction behaviors is a critical problem in modern e-commerce systems, particularly in scenarios where user intent evolves rapidly. Large Language Models (LLMs) offer strong semantic reasoning capabilities and have recently been adopted to enhance training data construction for next-query prediction. However, due to resource constraints on mobile devices, existing applications are deployed on cloud servers, resulting in high inference costs. In this paper, we propose \textbf{RecGPT-Mobile}, a framework that designs a lightweight LLM-based intent understanding agent to improve recommendation quality in mobile e-commerce scenarios. By deploying LLM directly on mobile devices, our approach can capture the evolving interests of users more quickly and adjust the recommendation results in real time. Extensive offline analyzes and online experiments demonstrate that our method significantly improves the accuracy of recommendation results, laying a practical path for LLM deployment in production-scale recommendation systems on mobile devices, as well as a scalable solution for integrating LLMs into real-world next-query prediction systems.
\end{abstract}




\maketitle

\section{Introduction}

The exponential growth of digital content have made personalized recommendation systems indispensable in modern applications. Although traditional cloud-based recommendation systems \cite{covington2016deep, guo2017deepfm, zhou2018deep} can utilize user behaviors effectively, it still struggles to meet the requirements for low-latency real-time inference in the mobile environment. Besides, centralized architectures introduce unavoidable communication delays between edge devices and remote servers and impede the perception of rapidly changing user intents. 

By building feature centers and collecting user behaviors locally, device-level recommendation systems \cite{gong2020edgerec, gong2022real, yin2025device, gu2022device}  make the results more real-time. However, deploying sophisticated model directly on resource-constrained devices presents significant challenges: LLMs \cite{bai2023qwen,achiam2023gpt,liu2024deepseek,yu2026deepresearchdeepresearch} are typically too massive in size and computationally intensive for direct mobile deployment. Recent advances in model compression techniques such as quantization \cite{frantar2023gptqaccurateposttrainingquantization}, pruning \cite{han2015learningweightsconnectionsefficient}, knowledge distillation \cite{hinton2015distillingknowledgeneuralnetwork} have emerged to bridge this gap, making it possible to deploy complex models on mobile devices.

To address these limitations, we propose RecGPT-Mobile, a novel device-level framework that deploys a lightweight LLM as an intent agent directly on the user’s mobile device.
In summary, our contributions are as follows:
\begin{itemize}
    \item  To our knowledge, this work presents the first implementation of an LLM-based recommendation system on mobile devices. We apply preference-based optimization and lightweight model compression to enable efficient deployment of LLM-based retrieval systems.
    \item  We design an intent agent that translates implicit user behaviors into explicit intent queries, as well as a frequency control mechanism to reduce redundant inference on the client device and improve resource utilization.
    \item We conducted extensive experiments on Mobile Taobao, involving millions of real users and diverse shopping sessions. The results show that RecGPT-Mobile significantly improves the relevance and interpretability of recommendations. 
\end{itemize}

\begin{figure}[t]
  \centering
  \includegraphics[width=\linewidth]{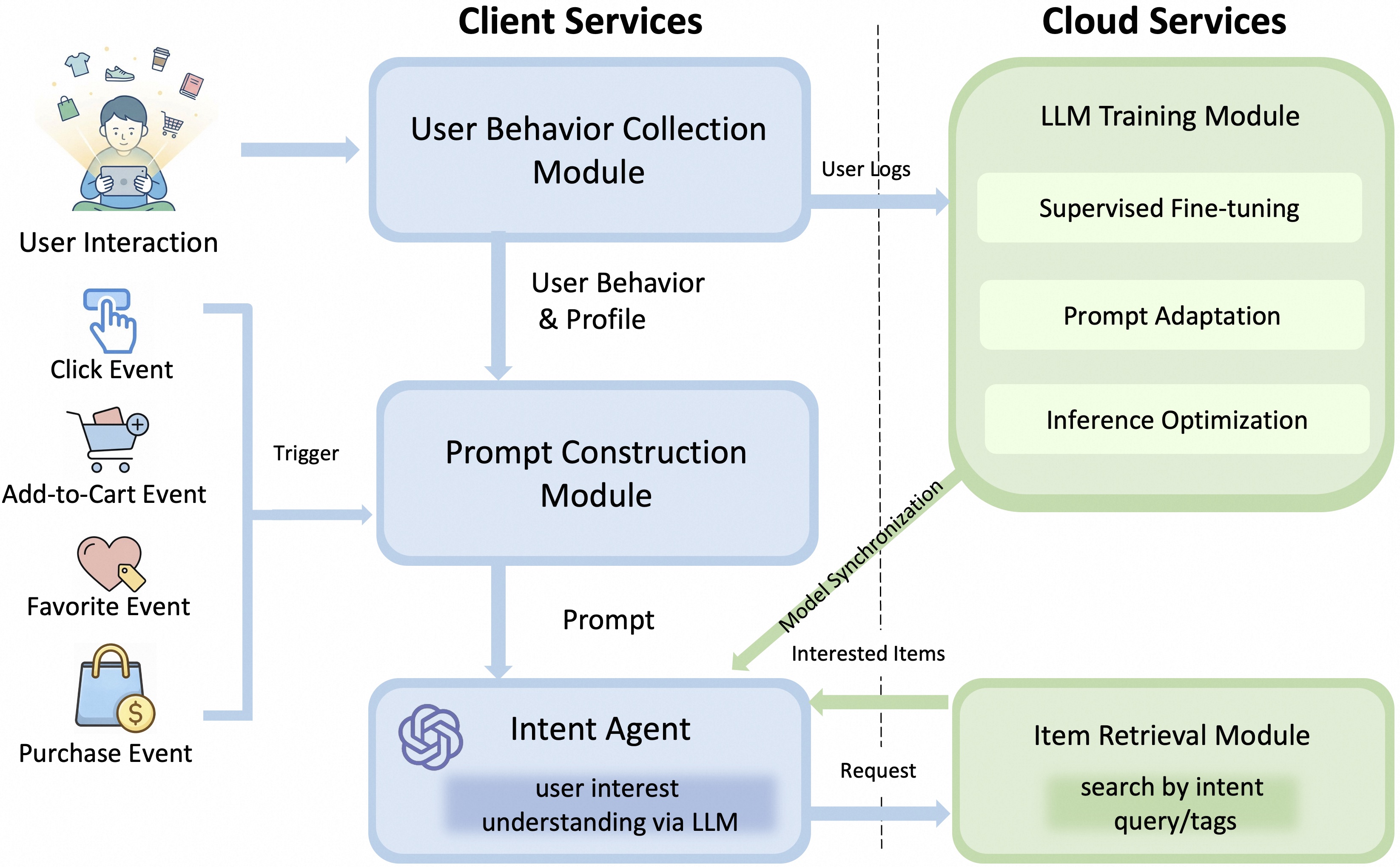}
  \caption{Framework of RecGPT-Mobile.}
  \label{Framework}
\end{figure}












\section{Related Work}

\textbf{On-device recommendation Systems.} EdgeRec \cite{gong2020edgerec} was the first to implement ranking models directly on mobile devices to reduce signal latency. Gong et al. \cite{gong2022real} implemented a real-time recommendation framework in the Kuaishou app, which processes user feedback locally on the device. To tackle the limitations of delayed processing in cloud-based re-ranking, DIR \cite{xi2023device} directly integrates re-ranking framework on mobile devices. 

\textbf{Cloud-based LLMs for recommendation Systems.} HSTU \cite{zhai2024actions} reformulated recommendation as a sequential transduction task within a generative framework. OneRec \cite{deng2025onerec} is a unified end-to-end generative framework that replaces the traditional cascaded strategy. RecGPT \cite{yi2025recgpttechnicalreport} is a LLM-based production-scale recommendation framework that replaces log-fitting methods with intent-centric reasoning. It integrates user interest mining, item tagging, retrieval, and explanation generation into a closed loop.

\textbf{Mobile LLMs.} Deploying LLMs on resource-constrained edge devices \cite{xu2024ondevicelanguagemodelscomprehensive,wang2025empowering,lu2025demystifying,wang2024mnnllm} is crucial for low latency and data locality, necessitating advances in architecture and compression to reduce memory use. Collaborative edge-cloud deployment and hardware-aware optimizations, such as leveraging mobile GPUs and NPUs, help balance resource usage and inference speed.

\section{Methodology}
\subsection{System Overview}
Fig~\ref{Framework} shows the overall architecture of RecGPT-Mobile framework. \textbf{User Behavior Collection Module} functions as a local repository to collect and cache user behavior data. Triggered by specific high-intent actions, \textbf{Prompt Construction Module} synthesizes raw behavior and profile data into a structured prompt.  By processing the synthesized prompt, the \textbf{Intent Agent} transforms complex behavioral signals into a clear user intent query. Upon receiving a request from the intent agent, \textbf{Item Retrieval Module} executes a search based on the generated intent query, and returns a set of interested items back to the client side. The \textbf{LLM Training Module} employs supervised fine-tuning, prompt adaptation and inference optimization to ensure the model remains both adaptive and accurate. In the following sections, we focus on the design and implementation of the LLM Training Module.

\subsection{Preliminaries}
We focus on the next-query prediction task given heterogeneous user behavior sequences. Formally, let a user behavior sequence be:
\begin{equation}
\small
\mathcal{B} = \{(i_1, a_1, t_1), (i_2, a_2, t_2), \dots, (i_T, a_T, t_T)\}, 
\end{equation}
where $i_T$ denotes the item interacted with at timestamp $T$, $a_T \in \{\text{click}, \text{cart}, \text{favorite}, \text{purchase}\}$ denotes the action type, and $t_T$ denotes the action time. Given the observed behavior sequence $\mathcal{B}$, the task is to predict the next potential search query $q \in \mathcal{Q}$, where $\mathcal{Q}$ is the space of search queries. The objective can be expressed as:
\begin{equation}
\small
q^* = \arg\max_{q \in \mathcal{Q}} P(q \mid \mathcal{B}) \text { s.t. Resource Constrains},
\end{equation}
where $P(q \mid \mathcal{B})$ captures the conditional probability of a search query given the user’s recent behaviors.

\begin{table}[t]
\centering
\caption{Training Sample Construction and Composition}
\label{tab:sample_sources}

\renewcommand{\arraystretch}{1.15}
\setlength{\tabcolsep}{3pt}

\begin{tabular}{p{2.6cm} p{3.6cm} p{0.8cm}}
\toprule
\textbf{Sample Type} & \textbf{Data Source} & \textbf{Ratio} \\
\midrule

Behavior-driven
& Purchase \& search logs
& 60\% \\

\addlinespace

Co-purchase
& Co-purchase item matrix
& 20\% \\

\addlinespace

LLM-based
& GPT-based  rewriting
& 15\% \\

\addlinespace

Human-annotated
& Manually annotated data
& 5\% \\

\bottomrule
\end{tabular}
\end{table}

\subsection{Supervised Fine-tuning}
To effectively model next-query prediction from heterogeneous user behavior sequences, we construct training data from multiple complementary sources, each capturing different aspects of user intent. As shown in Table~\ref{tab:sample_sources}, our training set consists of four types of samples: \textbf{behavior-driven samples}, extracted from purchase logs and search logs by linking purchased items to post-purchase queries and inferring complementary relations; \textbf{co-purchase relation samples}, automatically generated from the item-level co-purchase matrix to capture complementary signals from purchase co-occurrence; \textbf{LLM-based augmentation}, which rewrites rule-based samples with LLM to increase linguistic diversity while preserving semantics; and \textbf{human-annotated samples}, a small manually reviewed set used for quality calibration and reliable evaluation. \textbf{Prompt 1} below is used for supervised fine-tuning of the intent agent.

\begin{tcolorbox}[
  colback=white,
  colframe=black,
  boxrule=0.8pt,
  arc=4pt,
  left=6pt,
  right=6pt,
  top=6pt,
  bottom=6pt
]
\textbf{Prompt 1: Next-query Prediction Prompt.}

\vspace{0.5em}

\textbf{Input:}
Given a timestamp $t$ and a user behavior sequence
$\mathcal{B} = \{(i_1, a_1, t_1), (i_2, a_2, t_2), \ldots, (i_n, a_n, t_n)\}$, please infer the user's next search intent, and generate the most likely search query that reflects the user's latent requirement.

\vspace{0.5em}

\textbf{Output:}
<Predicted search query>.
\end{tcolorbox}

\subsection{Adaptive Prompt Construction}

\begin{algorithm}[t]
\caption{Adaptive Prompt Construction with Template \& Structural Adaptation}
\label{alg:adaptive_prompt}
\begin{algorithmic}[1]
\State \textbf{Input:} Behavior sequence $\mathcal{B}=\{(i_t,a_t,t_t)\}_{t=1}^{T}$; scenario $s$;
template pool $\mathcal{T}$; component set $\mathcal{C}$; scorer $M_{\mathrm{score}}$;
on-device budget $C_{\max}$ (latency/memory/token).
\State \textbf{Output:} Adaptive prompt $P^\ast$.
\Statex
\State \textbf{Stage 1: Feature Extraction}
\State Compute behavior features $\Phi \gets \Phi(\mathcal{B}) = [\Phi_{\mathrm{act}}, \Phi_{\mathrm{rec}}, \Phi_{\mathrm{div}}, \Phi_{\mathrm{freq}}]$,
where $\Phi_{\mathrm{act}}$ represents action type, $\Phi_{\mathrm{rec}}$ encodes recency, $\Phi_{\mathrm{div}}$ denotes diversity, and $\Phi_{\mathrm{freq}}$ is the action frequency.

\State \textbf{Stage 2: Template-level Adaptation}
\ForAll{$T_k \in \mathcal{T}_s$}
    \State $\alpha_k \gets M_{\mathrm{score}}(T_k, \Phi, s)$
\EndFor
\State $p_k \gets \exp(\beta \alpha_k)\,/\,\sum_j \exp(\beta \alpha_j)$
\State $T^\ast \gets \arg\max_{T_k \in \mathcal{T}_s} p_k$
\State Initialize prompt $P \gets T^\ast$

\State \textbf{Stage 3: Structural-level Adaptation}
\ForAll{$c \in \mathcal{C}$}
    \State $\Delta(c) \gets M_{\mathrm{score}}(P \oplus c, \Phi, s) - M_{\mathrm{score}}(P, \Phi, s)$
    \If{$\Delta(c) > \tau$ \textbf{and} $\mathrm{Cost}(P \oplus c) \le C_{\max}$}
        \State $P \gets P \oplus c$
    \EndIf
\EndFor

\State \textbf{Stage 4: Budget Enforcement \& Finalization}
\State $P^\ast \gets \arg\max_{P' \subseteq P} M_{\mathrm{score}}(P', \Phi, s)$ \textbf{s.t.} $\mathrm{Cost}(P') \le C_{\max}$
\State Instantiate behavior tokens from $\mathcal{B}$ into $P^\ast$
\State \Return $P^\ast$
\end{algorithmic}
\end{algorithm}

Algorithm~\ref{alg:adaptive_prompt} describes the adaptive prompt construction procedure for next-query prediction under on-device constraints.
Given a user behavior sequence $\mathcal{B}$ and the scenario context $s$, the algorithm first summarizes heterogeneous user interactions into a compact behavior feature vector $\Phi(\mathcal{B})$, which captures action distribution, temporal recency, semantic diversity, and interaction frequency.

Based on the extracted features, the algorithm performs template-level adaptation by scoring candidate prompt templates using a lightweight scoring model and selecting the most suitable template conditioned on both the behavior characteristics and scenario context. After the template is selected, structural-level adaptation is applied to incrementally refine the prompt structure. Specifically, candidate structural components are evaluated according to their marginal utility gain, and only those that contribute positively while satisfying the on-device budget constraints are incorporated into the prompt. Finally, a budget-aware pruning step is conducted to ensure that the constructed prompt maximizes utility under strict latency, memory, and token constraints. The resulting prompt is instantiated with concrete behavior tokens and used as the input to the on-device language model for next-query prediction.

\begin{figure}[t]
  \centering
  \includegraphics[width=\linewidth]{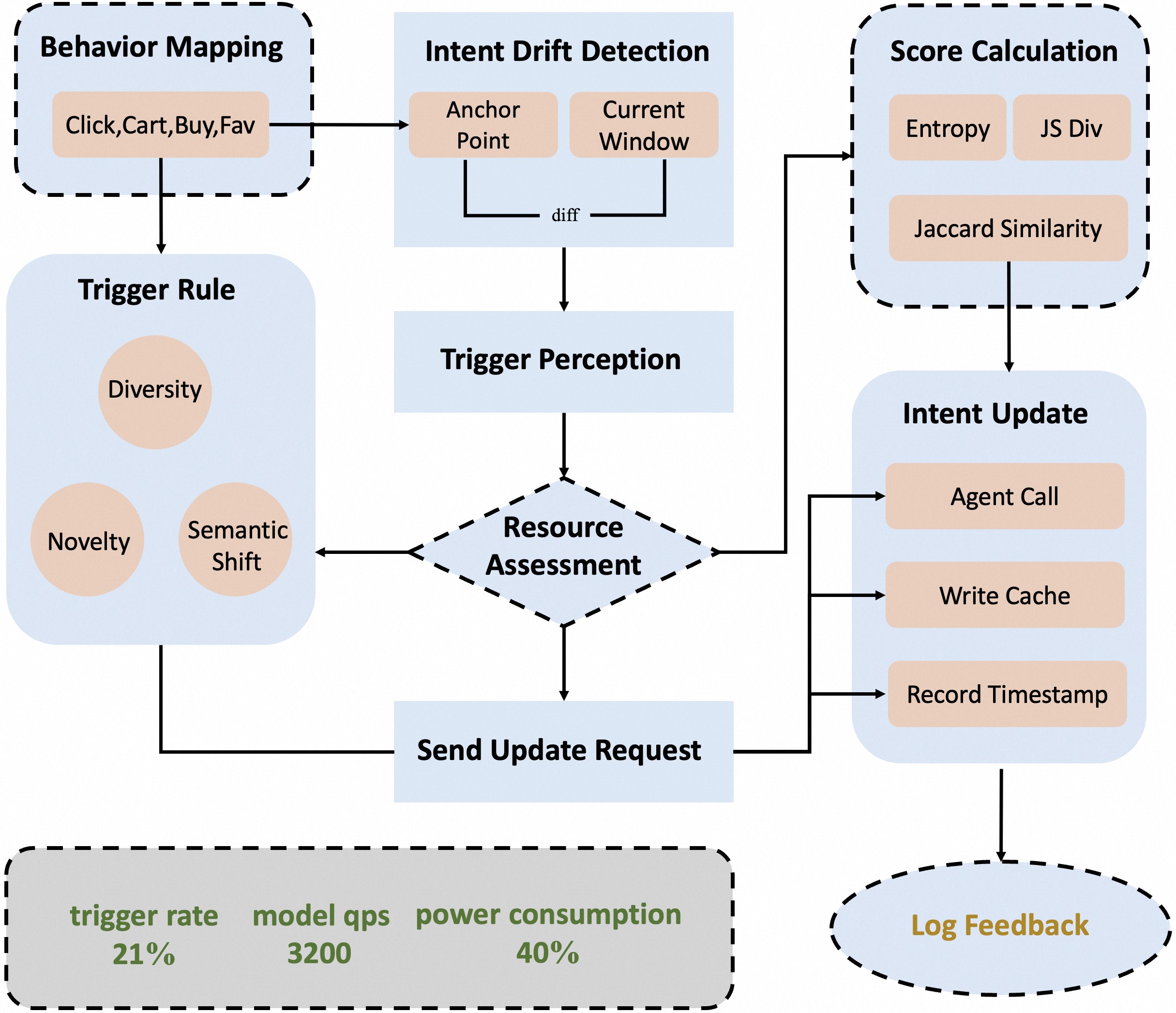}
  \caption{Mobile Intent Agent Trigger Pipeline.}
  \label{Pipeline}
\end{figure}

\subsection{Mobile Device Inference Optimization}
The core objective of the Intent Agent is to update semantic representation when the user's intent changes, while avoiding frequent computation when behavior is stable, thus achieving a balance between on-device running performance and recommendation accuracy. To this end, we designed the intent agent trigger pipeline, as shown in Fig~\ref{Pipeline}.

Given a user behavior sequence $\mathcal{B}$ within a sliding window, each interaction is mapped to a discrete semantic tag (e.g., category, brand, or intent type). This yields a normalized tag distribution $P_{\mathcal{B}}$ over the current window. Let $P_{\mathcal{B}}^{(t)}$ and $P_{\mathcal{B}}^{(t-1)}$ denote the tag distributions at the current step and the previous trigger point, respectively. We quantify intent drift from three complementary perspectives. First, we measure the change in uncertainty of user intent using entropy:
\begin{equation}
\small
H(P_{\mathcal{B}}) = - \sum_{k} P_{\mathcal{B}}(k)\log P_{\mathcal{B}}(k).
\end{equation}
The absolute entropy difference
\begin{equation}
\small
\Delta H = \big| H(P_{\mathcal{B}}^{(t)}) - H(P_{\mathcal{B}}^{(t-1)}) \big|
\end{equation}
captures whether the user intent becomes more focused or more exploratory. Second, we assess semantic overlap between consecutive behavior windows using Jaccard similarity:
\begin{equation}
\small
JA(\mathcal{Z}^{(t)}, \mathcal{Z}^{(t-1)}) =
\frac{|\mathcal{Z}^{(t)} \cap \mathcal{Z}^{(t-1)}|}
{|\mathcal{Z}^{(t)} \cup \mathcal{Z}^{(t-1)}|},
\end{equation}
where $\mathcal{Z}^{(t)}$ denotes the set of observed tags. A lower Jaccard score indicates a larger shift in semantic focus. Finally, we explicitly model distributional drift using Jensen–Shannon (JS) divergence:
\begin{equation}
\small
\mathrm{JS}\!\left(P_{\mathcal{B}}^{(t)}, P_{\mathcal{B}}^{(t-1)}\right)
=
\frac{1}{2}\mathrm{KL}\!\left(P_{\mathcal{B}}^{(t)} \,\middle\|\, M\right)
+
\frac{1}{2}\mathrm{KL}\!\left(P_{\mathcal{B}}^{(t-1)} \,\middle\|\, M\right),
\end{equation}
where $M = \frac{1}{2}\left(P_{\mathcal{B}}^{(t)} + P_{\mathcal{B}}^{(t-1)}\right)$. We fuse the above signals into a single intent drift score:
\begin{equation}
\small
\Delta_{\text{intent}} =
\lambda_1 \cdot \Delta H
+
\lambda_2 \cdot \big(1 - JA(\mathcal{Z}^{(t)}, \mathcal{Z}^{(t-1)})\big)
+
\lambda_3 \cdot \mathrm{JS}\!\left(P_{\mathcal{B}}^{(t)}, P_{\mathcal{B}}^{(t-1)}\right),
\end{equation}
where $\lambda_1, \lambda_2, \lambda_3 \ge 0$ and $\lambda_1 + \lambda_2 + \lambda_3 = 1$ control the relative importance of uncertainty change, semantic overlap, and distributional drift. The LLM is triggered to update the prompt and generate a new prediction only if
\begin{equation}
\small
\Delta_{\text{intent}} > \tau,
\end{equation}
with $\tau$ being a predefined threshold.

\section{Experiments}
\subsection{Offline Evaluation}
\textbf{Prompt 2} is designed to evaluate and decompose generation quality into three complementary dimensions: semantic relevance ($S_{\text{sem}}$), logical consistency ($S_{\text{logic}}$), and linguistic style ($S_{\text{style}}$). The final score is computed as a weighted aggregation of the three sub-scores, providing a holistic assessment of generation quality.

\begin{tcolorbox}[
  colback=white,
  colframe=black,
  boxrule=0.8pt,
  arc=4pt,
  left=6pt,
  right=6pt,
  top=6pt,
  bottom=6pt
]
\textbf{Prompt 2: LLM-based Evaluation Prompt.}

\vspace{0.6em}

\textbf{Input:}
Given a user behavior sequence
$\mathcal{B} = \{(i_1,a_1,t_1), (i_2,a_2,t_2), \ldots, (i_n,a_n,t_n)\}$,
and a candidate search query $q$ generated for next-query prediction,
please evaluate the quality of $q$ from the following three aspects:

\begin{itemize}
  \item \textbf{Semantic Consistency}: whether $q$ is semantically aligned with the latent intent implied by $\mathcal{B}$.
  \item \textbf{Logical Coherence}: whether $q$ reflects a reasonable intent transition given the behavior sequence.
  \item \textbf{Expression Quality}: whether $q$ avoids trivial template reuse and is expressed in a natural manner.
\end{itemize}

\vspace{0.6em}

\textbf{Output:}
Three normalized scores
$S_{\text{sem}}, S_{\text{logic}}, S_{\text{style}} \in [0,1]$
\end{tcolorbox}

Table~\ref{tab:auto_eval} reports the automatic evaluation results across different evaluation model sizes and deployment settings. We use Qwen3-0.6B as the base model and compare to LoRA-adapted models\cite{hu2022lora}, and quantized LoRA models \cite{dettmers2022gpt3} under the same evaluation protocol. Overall, LoRA models achieve the highest scores across most dimensions, indicating the effectiveness of lightweight fine-tuning. Notably, quantized LoRA models consistently preserve a large portion of the semantic, logical, and stylistic quality of their full-precision counterparts, with only marginal degradation in total scores. This shows that our approach remains robust under resource-constrained deployment scenarios, validating the practicality of on-device inference with minimal quality loss.
\begin{table}[t]
\centering
\caption{LLM-based Automatic Evaluation Results}
\label{tab:auto_eval}
\small
\renewcommand{\arraystretch}{1.1}

\begin{tabular}{
p{1.6cm}  
p{1.6cm}  
c c c c   
}
\toprule
\textbf{Eval. Model} & \textbf{Target}
& $S_{\text{sem}}$ & $S_{\text{logic}}$ & $S_{\text{style}}$ & \textbf{Total} \\
\midrule

\multirow{3}{=}{Qwen3-4B}
& Base  & 0.752 & 0.621 & 0.657 & 0.677 \\
& \textbf{LoRA}  & \textbf{0.885} & \textbf{0.792} & \textbf{0.811} & \textbf{0.829} \\
& LoRA+Quant
& 0.844 & 0.754 & 0.785 & 0.794 \\

\midrule

\multirow{3}{=}{Qwen3-8B}
& Base  & 0.657 & 0.554 & 0.627 & 0.613 \\
& \textbf{LoRA}  & \textbf{0.807} & \textbf{0.755} & \textbf{0.786} & \textbf{0.783} \\
& LoRA+Quant
& 0.780 & 0.746 & 0.754 & 0.760 \\

\midrule

\multirow{3}{=}{Qwen3-30B}
& Base  & 0.671 & 0.654 & 0.663 & 0.654 \\
& \textbf{LoRA}  & \textbf{0.839} & \textbf{0.797} & \textbf{0.787} & \textbf{0.808} \\
& LoRA+Quant
& 0.812 & 0.774 & 0.762 & 0.780 \\

\bottomrule
\end{tabular}
\end{table}

\subsection{Online A/B Testing}
\begin{figure}[t]
  \centering
  \includegraphics[width=\linewidth]{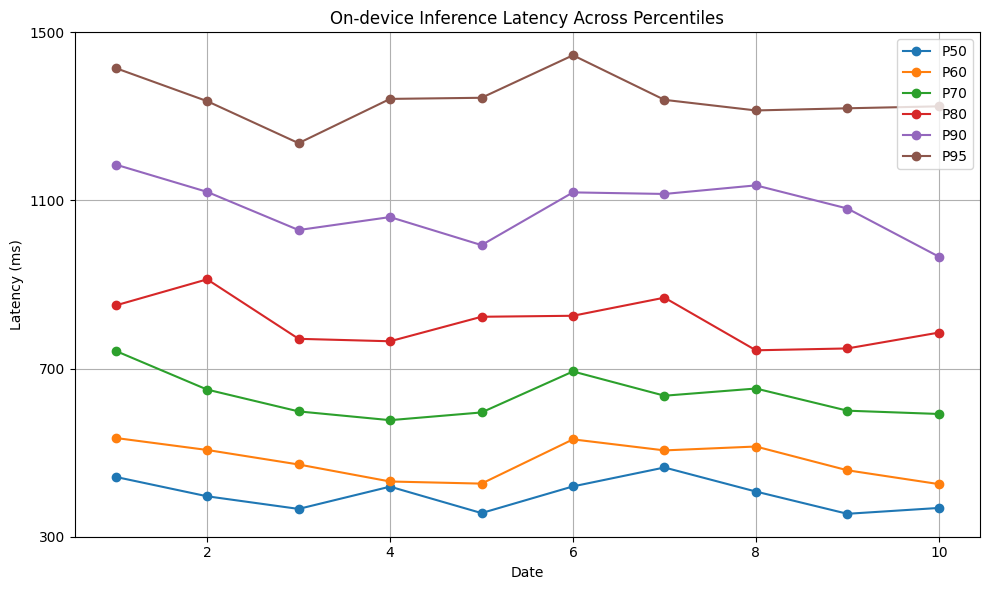}
  \caption{Running latency on real-world mobile devices under different percentile}
  \label{latency}
\end{figure}

We conducted online experiments in four scenarios of mobile Taobao during a one-month testing, covering tens of millions of users. Considering storage limitations, RecGPT-Mobile adopts Qwen3-0.6B-Quant as mobile deployment model, with hyperparameter settings of $\lambda_1=0.4, \lambda_2=0.3, \lambda_3=0.3, \tau=0.8$, which were determined via experimental heuristic search. As shown in Table~\ref{online}, RecGPT-Mobile achieves a definitely significant improvement across all four feed scenarios, contributing 1.8\% CLICK, 2.7\% PAY and 2.5\% GMV promotions on average.
Fig~\ref{latency} presents the on-device inference latency across multiple percentiles on different dates. While higher percentiles exhibit increased latency as expected, the overall temporal trends remain consistent across P50 to P95. This indicates that the model maintains stable execution behavior even under amplified tail-latency conditions. Moreover, the moderate divergence among percentile curves reflects realistic device-level variance rather than systematic performance degradation, suggesting that the proposed deployment is robust to runtime fluctuations commonly encountered in real-world client environments.

\begin{table}[h!]
\centering
\caption{Results of online experiment.}
\label{tab:ablation}
\small
\setlength{\tabcolsep}{4pt}
\begin{tabular}{lccc}
\toprule
\textbf{Scenario} & \textbf{CLICK} & \textbf{PAY} & \textbf{GMV} \\
\midrule
Payment Success Page
& +1.3\% & +2.3\% & +2.5\% \\

Shipment Tracking Page
& +2.4\% & +2.9\% & +3.0\% \\

Shopping Cart Page
& +2.5\% & +2.7\% &+2.9\% \\

Order List Page
& +0.8\% & +1.8\% & +1.8\% \\

\midrule

Average
& +1.8\% & +2.7\% & +2.5\% \\

\bottomrule
\label{online}
\end{tabular}
\end{table}

\section{Conclusion}
This paper presents an on-device framework for next-query intent prediction from user behavior sequences, leveraging large language models with adaptive prompt construction under strict resource constraints. By dynamically adjusting prompt templates and structures, and incorporating a mobile device trigger optimization scheme, the proposed approach enables effective training and robust inference without extensive human supervision. Experimental results show that the model maintains strong intent understanding ability while preserving stable runtime performance across various tail-latency conditions, demonstrating its practicality for deploying recommendation systems on mobile devices.

\clearpage

\section*{Presenter Biography}
Bin Zhang is an algorithm expert at Taobao, focusing on researching and applying large language models in recommendation systems that enhance user experience through a deeper understanding of user intent. During his work at Taobao, he has applied large language models to develop innovative methods for user intent understanding in recommendations.
\bibliographystyle{ACM-Reference-Format}
\bibliography{sample-base}


\end{document}